\documentclass[preprint,12pt]{elsarticle}
\journal{Journal of Magnetism and Magnetic Materials}
\usepackage{amsmath,amssymb}
\usepackage{bm}
\PassOptionsToPackage{hyphens}{url}
\usepackage[hyphens]{url}
\usepackage{hyperref}
\hypersetup{colorlinks = true}
\usepackage[anythingbreaks]{breakurl}
\usepackage{color}

\begin{document}

\begin{frontmatter}

  \title{Substantial reduction of write-error rate for voltage-controlled magnetoresistive random access memory by in-plane demagnetizing field and voltage-induced negative out-of-plane anisotropy field}

  \author{Rie Matsumoto}
  \ead{rie-matsumoto@aist.go.jp}
  \author{Shiniji Yuasa}
  \author{Hiroshi Imamura}
  \ead{h-imamura@aist.go.jp}
  \address{National Institute of Advanced Industrial Science and Technology (AIST), Tsukuba, Ibaraki 305-8568, Japan}

  \begin{abstract}
Voltage-controlled magnetoresistive random access memory (VC-MRAM) based on voltage-induced dynamic switching in magnetic tunnel junctions (MTJs) is a promising ultimate non-volatile memory with ultralow power consumption.
However, the dynamic switching in a conventional MTJ is accompanied by a relatively high write error rate (WER), hindering the reliable operation of VC-MRAM. Here, we propose a reliable writing scheme using the in-plane demagnetizing field (IDF) and voltage-induced negative out-of-plane anisotropy field (NOAF). 
Numerical simulations based on macrospin model demonstrate that the voltage-induced NOAF modifies the switching dynamics and increases the torque due to the IDF, thereby reducing the switching time. The IDF and voltage-induced NOAF also reduce the mean energy difference between the magnetization direction at the end of the pulse and the equilibrium direction. As a result, an appropriate combination of the IDF and voltage-induced NOAF reduces the WER by one order of magnitude compared with that of the dynamic switching in a conventional MTJ.
  \end{abstract}


\end{frontmatter}

\section{Introduction}
\label{introduction}
\begin{figure}[t]
  \centerline{
    \includegraphics [width=0.8\columnwidth] {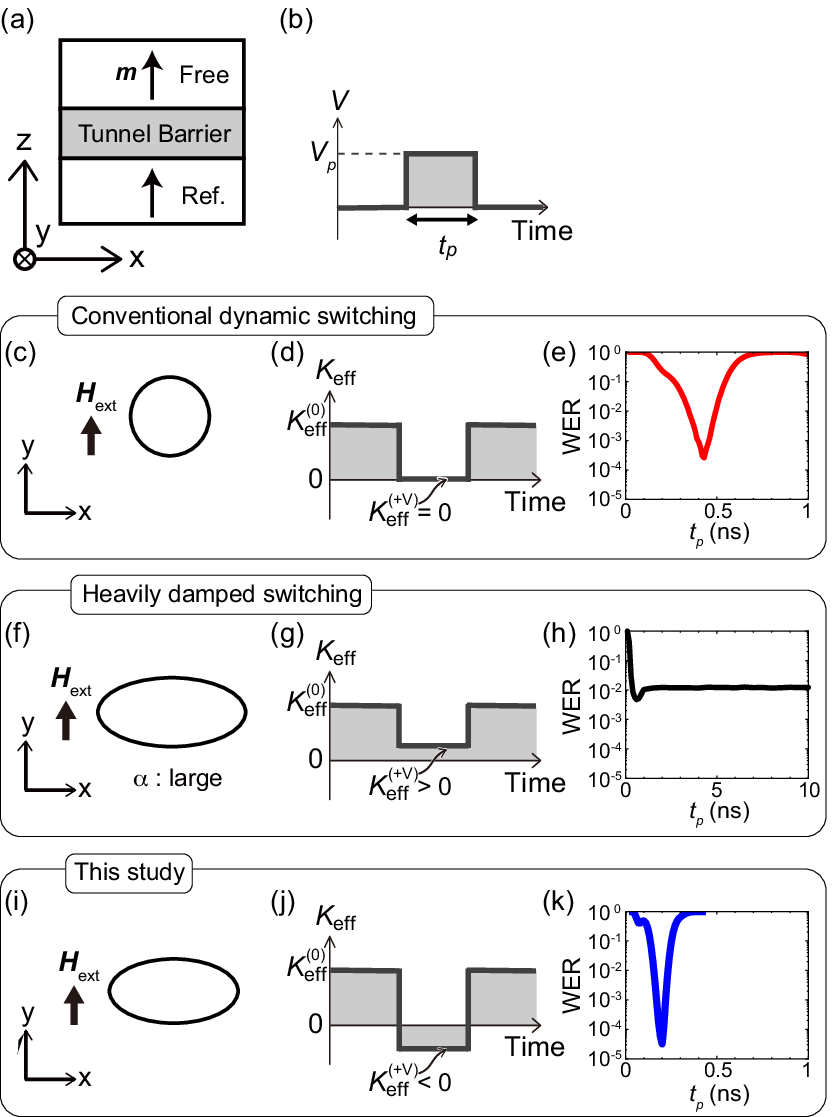}
  }
  \caption{
    \label{fig:fig1}
    (a) Cross-section of a magnetic tunnel junction (MTJ) with a perpendicularly-magnetized free layer and a reference (Ref.) layer, and the Cartesian coordinates defined as $(x, y, z)$.
    (b) Shape of the voltage pulse. The amplitude and duration of the pulse are $V_{p}$ (positive value) and $t_{p}$, respectively.
    (c) Top view of the MTJ and the direction of the external magnetic field, $\bm{H}_{\rm ext}$, used in the conventional precessional switching (CPS) scheme. A small Gilbert damping constant, $\alpha$, is preferred.
    (d) Temporal variation of the effective out-of-plane anisotropy constant, $K_{\rm eff}$, used in the CPS scheme.
    (e) An example of $t_{p}$ dependence of the write error rate (WER) obtained by the CPS scheme.
    (f) Top view of the MTJ and the direction of $\bm{H}_{\rm ext}$ used in the heavily damped switching (HDS) scheme. A large $\alpha$ is preferred.
    (g) Temporal variation of $K_{\rm eff}$ used in the HDS scheme.
    (h) Typical example of $t_{p}$ dependence of WER obtained by the HDS scheme.
    (i) Top view of the MTJ and the direction of $\bm{H}_{\rm ext}$ used in the switching scheme proposed in this study. A small $\alpha$ is preferred.
    (j) Temporal variation of $K_{\rm eff}$ used in this study.
    (k) Typical example of $t_{p}$ dependence of WER obtained in this study.
    In (d), (g), (j), $K_{\rm eff}^{(0)}$ and $K_{\rm eff}^{(+V)}$ denote $K_{\rm eff}$ at voltage amplitudes of $V=0$ and $V=V_{p}$, respectively. 
  }
\end{figure}

Voltage-controlled magnetoresistive random-access memory (VC-MRAM)
\cite{weisheit_electric_2007,maruyama_large_2009, duan_surface_2008,nakamura_giant_2009, tsujikawa_finite_2009,endo_electric-field_2010, shiota_induction_2012, shiota_pulse_2012,kanai_electric_2012,shiota_evaluation_2016, grezes_ultra-low_2016, shiota_reduction_2017,yamamoto_thermally_2018, yamamoto_improvement_2019} has attracted considerable attention 
as an emerging ultralow-power non-volatile memory. 
The writing scheme of VC-MRAM is based on the voltage control of magnetic anisotropy (VCMA) 
in a magnetic tunnel junction (MTJ) \cite{yuasa_giant_2004, parkin_giant_2004, djayaprawira_230_2005} 
with perpendicular magnetization (see Fig. \ref{fig:fig1}(a)). 
Without applied voltage, the magnetization in the FL is aligned nearly in the out-of-plane direction due to the perpendicular magnetic anisotropy (PMA) at the interface. Application of the voltage pulse (see Fig. \ref{fig:fig1}(b)) reduces the PMA through the VCMA effect \cite{weisheit_electric_2007, maruyama_large_2009, duan_surface_2008,nakamura_giant_2009, tsujikawa_finite_2009},
inducing the precession of the magnetization around the magnetic field applied in the in-plane direction \cite{davies_anomalously_2019}.

In the conventional dynamic switching scheme \cite{endo_electric-field_2010, shiota_induction_2012, shiota_pulse_2012,kanai_electric_2012,shiota_evaluation_2016, grezes_ultra-low_2016, shiota_reduction_2017,yamamoto_thermally_2018, yamamoto_improvement_2019}, 
the circular cylinder shaped MTJ nanopillar is used as shown Fig. \ref{fig:fig1}(c). The effective out-of-plane anisotropy constant, $K_{\rm eff}$, is reduced to zero by the voltage pulse, as shown in Fig. \ref{fig:fig1}(d), where $K_{\rm eff}^{(0)}$ and $K_{\rm eff}^{(+V)}$ denote the effective out-of-plane anisotropy at $V=0$ and $V=V_{p}$, respectively. 
The WER strongly depends on the the pulse duration, $t_{p}$, as shown in Fig. \ref{fig:fig1}(e). 
The WER is minimized when $t_{p}$ is approximately the half of the precession period. 
The VC-MRAM has a clear advantage that its energy consumption to write a bit  is 
one hundred times smaller than that of the spin-transfer-torque (STT)-MRAM
\cite{kanai_electric-field-induced_2016,grezes_ultra-low_2016}. 
However, the WER of the VC-MRAM ($\gtrsim 10^{-6}$)   
 \cite{yamamoto_improvement_2019} is considerably higher than that of the STT-MRAM ($\sim 10^{-11}$)
 \cite{apalkov_magnetoresistive_2016}.
To reduce the WER of VC-MRAM, very precise control of the pulse duration is necessary. 
Notably, such precise control of the pulse duration is difficult in large-scale integrated circuits due to circuit delay.

Another switching scheme of VC-MRAM called heavily damped switching \cite{matsumoto_voltage-induced_2019,matsumoto_heavily_2022} is schematically shown in Fig. \ref{fig:fig1}(f). The in-plane magnetic field is applied parallel to the minor axis of the elliptical MTJ. The Gilbert damping constant, $\alpha$, is assumed to be large enough to suppress the precession back to the initial direction. The effective out-of-plane anisotropy constant is reduced by the voltage pulse but remains positive as shown in Fig. \ref{fig:fig1}(g). 
Although the minimum WER of the heavily damped switching is higher than that of the conventional dynamic switching 
due to the large $\alpha$, the heavily damped switching has an advantage that the WER is insensitive 
to the pulse duration 
as shown in Fig. \ref{fig:fig1}(h). Therefore, the precise control of the pulse duration is unnecessary. 
The heavily damped switching is suited for some special applications 
such as error-tolerant machine learning for image recognition and object detection \cite{yeoh_jsap_2022}.

For cache applications, 
it is necessary to develop a reliable writing scheme, whose WER is very low. 
In this paper, we propose a new reliable switching scheme using the in-plane demagnetizing field (IDF) and voltage-induced negative out-of-plane anisotropy field (NOAF) (Figs. \ref{fig:fig1}(i) and (j)). 
We calculated the WER by solving the Langevin equation of the macrospin model,
demonstrating that the WER is reduced by one order of magnitude (Fig. \ref{fig:fig1}(k)) 
compared with that of the dynamic switching in a conventional MTJ (Fig. \ref{fig:fig1}(e)). 
The mechanism of the reduction of the WER is discussed by analyzing the switching dynamics, torques, and distribution of the magnetizations at the end of the pulse.

\section{Model and method}
We consider an elliptical cylinder shaped MTJ nanopillar, shown in Figs. \ref{fig:fig1}(a) and (i). The lateral size of the nanopillar is assumed to be so small that the magnetization dynamics can be described by the macrospin model. The direction of magnetization in the FL is represented by the unit vector ${\bm m} = (m_{x}$, $m_{y}$, $m_{z}) = (\sin \theta \cos \phi$, $\sin \theta \sin \phi$, $\cos \theta$),
where $\theta$ and $\phi$ are the polar and azimuthal angles, respectively.
The $x$-axis is parallel to the major axis of the ellipse. The external in-plane magnetic field, ${\bm H}_{\rm ext}$, is applied in the positive $y$-direction.
The magnetization in the reference layer is fixed to align with the
positive $z$-direction.

The energy density of the FL is given by
\cite{stiles_spin-transfer_2006}
\begin{eqnarray}
  \label{eq:E}
  {\cal E} (m_{x}, m_{y}, m_{z})
  =
  &
  \frac{1}{2} \mu_{0} M_{s}^{2}
  ( N_{x} m_{x}^{2} + N_{y} m_{y}^{2} + N_{z} m_{z}^{2} )
  \nonumber\\
  &
  + K_{u} (1-m_{z}^{2})
  - \mu_{0} M_{s} {\bm m} \cdot \bm{H}_{\rm ext},
\end{eqnarray}
where, the demagnetization coefficients $N_{x}$, $N_{y}$, and $N_{z}$ are assumed to satisfy $N_{z} \gg N_{y} > N_{x}$, $\mu_{0}$ is the vacuum permeability, $M_{s}$ is the saturation magnetization of the FL, and $\bm{H}_{\rm ext}=(0$, $H_{\rm ext}, 0)$ is the external in-plane magnetic field. Without loss of generality, we assume that $H_{\rm ext} > 0$. The index of the IDF, $H_{D}$ is given by $H_{D}=M_{s}(N_{y} - N_{x})$ \cite{matsumoto_critical_2016}. $K_{u}$ is the uniaxial out-of-plane anisotropy constant. The value of $K_{u}$ can be controlled by applying a bias voltage $V$ through the VCMA effect, as shown in Fig. \ref{fig:fig1}(j). Hereafter, $K_{\rm eff}$ represents the effective out-of-plane anisotropy constant defined by $K_{\rm eff} = K_{u} - (1/2) \mu_{0} M_{s}^{2} ( N_{z} - N_{x} )$, and $K_{\rm eff}^{\rm (+V)}$ indicates the value of $K_{\rm eff}$ during the voltage pulse.
Note that, when $H_{\rm ext} = 0$ and $K_{\rm eff}$ is negative (positive), the magnetization is relaxed to in-plane (perpendicular) state.
We call the field induced by the negative effective out-of-plane anisotropy as negative out-of-plane anisotropy field (NOAF).

The magnetization dynamics are simulated using
the following Langevin equation \cite{brown_thermal_1963}:
\begin{equation}
  \label{eq:Langevin}
  (1+ \alpha^{2}) \frac{{\rm d} {\bm m}}{{\rm d}t}
  = -\gamma_{0} {\bm m}\times
  \left\{
  \left({\bm H}_{\rm eff} + {\bm h}\right)
  +\alpha
  \left[
    {\bm m}\times\left({\bm H}_{\rm eff} + {\bm h}\right)
    \right]
  \right\},
\end{equation}
where $t$ is time, $\gamma_{0}=2.21\times10^{5}$ ${\rm rad}\cdot {\rm s}^{-1}\cdot({\rm A/m})^{-1}$ is the gyromagnetic ratio, and $\alpha$ is the
Gilbert damping constant. The thermal agitation field, ${\bm h}$, satisfies the following relations: $\langle h_{\iota}(t)\rangle=0$ and
$
  \langle
  h_{\iota}(t)h_{\kappa}(t')
  \rangle
  = \left[2\alpha k_{\rm B} T / \left( \gamma_{0} \mu_{0} M_{s} \Omega \right)\right]\delta_{\iota\kappa}\delta(t-t')
$, where $\langle \rangle$ represents the statistical mean, $\iota,\kappa=x,y,z$,
$k_{\rm B}$ is the Boltzmann constant, $T$ is the temperature, $\Omega$ is the volume of the FL volume, and $\delta_{\iota\kappa}$ is Kronecker's delta. The effective magnetic field, ${\bm H}_{\rm eff}$, is defined as
\begin{equation}
  \label{eq:Heff}
  {\bm H}_{\rm eff}= -\frac {1}{\mu_{0} M_{s}} \nabla {\cal E}.
\end{equation}

The following parameters are assumed in simulations. 
The Gilbert damping constant is set as $\alpha=0.1$, the saturation magnetization as $M_{s}= 955$ kA/m, 
the effective anisotropy constant at $V=0$ as $K_{\rm eff}^{(0)} = 110$ kJ/m$^{3}$, and temperature as $T=300$ K. Hereafter, the superscript ``(0)'' denotes any quantities obtained at $V=0$. The thickness and area of the FL are $d=1.1$ nm and $S=289\pi$ nm$^{2}$, respectively. The aspect ratio of the ellipse is assumed to be $R_{\rm asp}=3$. We also performed simulations for circular MTJ, $R_{asp}=1$, for comparison.

The initial state of the simulation is prepared
by relaxing the magnetization at $T=300$ K with $K_{\rm eff}=K_{\rm eff}^{(0)} (> 0)$ for 10 ns
from the equilibrium direction at $T=0$ K with $m_{z} > 0$
\cite{IMAMURA2022170012}. 
Then, the magnetization dynamics at $T=300$ K are calculated under a voltage pulse applied over a duration of $t_{p}$ (Fig. \ref{fig:fig1}(b)).
During the duration of the pulse, $K_{\rm eff}$ is reduced to $K_{\rm eff}^{\rm (+V)}$ through the VCMA effect (Fig. \ref{fig:fig1}(j)). 
After the pulse, the anisotropy constant is increased to the initial value of 
$K_{\rm eff} = K_{\rm eff}^{(0)}$, and there the magnetization is relaxed at $T=300$ K  for 10 ns.
The success or failure of switching is determined by the sign of $m_{z}$ after 10 ns of relaxation. The demagnetization coefficient of the FL is $N_{x}=0.01817$, $N_{y}=0.08445$, $N_{z}=0.89738$
\cite{beleggia_demagnetization_2005}, and the magnitude of the IDF is
$H_{D} (= 63.3$ kA/m) $ = 795$ Oe
\cite{matsumoto_critical_2016}.
The WERs are calculated from $10^{6}$ trials.

\section{Results}

\subsection{Upper and lower boundaries of $K_{\rm eff}^{\rm (+V)}$ for dynamic switching}

\begin{figure}
  \centerline{
    \includegraphics [width=1\columnwidth] {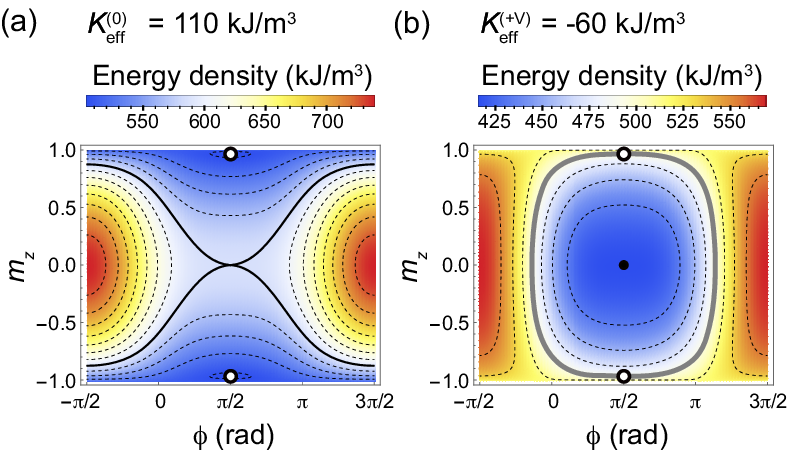}
  }
  \begin{center}
    \caption{
    \label{fig:fig2}
    (a) Contour plot of the energy density at $V=0$ on the $\phi$-$m_{z}$ plane. 
    The effective out-of-plane anisotropy constant is $K_{\rm eff}^{(0)}=110$ kJ/m$^{3}$. The solid black curve represents the energy contour crossing the saddle point at $(\phi,m_{z})=(\pi/2,0)$, i.e., the separatrix. The open circles indicate the equilibrium directions, $\bm{m}^{(0)}$.
    (b) Contour plot of the energy density during the pulse with $K_{\rm eff}^{\rm (+V)}=-60$ kJ/m$^{3}$. The open circles indicate the equilibrium directions, $\bm{m}^{(0)}$. The thick gray contour represents the contour with the same energy density as $\cal{E}$$(\bm{m}^{(0)})$, 
    corresponding to the trajectory of the magnetization precession with $\alpha=0$ at $T=0$ K.
    }
  \end{center}
\end{figure}

The equilibrium magnetization direction at $T=0$ and $V=0$ is obtained by minimizing the energy density at $V=0$, ${\cal E}^{(0)}$. Figure \ref{fig:fig2}(a) shows the contour plot of ${\cal{E}}^{(0)}$ at $H_{\rm ext}=800$ Oe on the $\phi$-$m_{z}$ plane, where the equilibrium directions, ${\bm m}^{(0)} = (m_{x}^{(0)}$, $m_{y}^{(0)}$, $m_{z}^{(0)})=(0$, 0.258, $\pm0.966)$, are indicated by the open circles. The application of a bias voltage reduces the anisotropy constants from $K_{\rm eff}^{(0)}$ to $K_{\rm eff}^{\rm (+V)}$ and destabilizes the initial state.
Under appropriate conditions of $K_{\rm eff}^{(0)}$ and $K_{\rm eff}^{\rm (+V)}$,
the precessional motion of magnetization around the effective magnetic field is induced \cite{matsumoto_voltage-induced_2018}. As shown in Fig. \ref{fig:fig2}(b), at $K_{\rm eff}^{\rm (+V)}=-60$ kJ/m$^{3}$, two equilibrium directions indicated by the open circles are connected by the closed thick gray contour. 
Therefore, using the precession when the pulse yielding $K_{\rm eff}^{\rm (+V)}=-60$ kJ/m$^{3}$ is applied for the half period of precession,
 the direction of the magnetization can be switched from one equilibrium direction to the other.

The analytical expression of the upper and lower boundaries of $K_{\rm eff}^{\rm (+V)}$ for dynamic switching is obtained by analyzing the stability of the equilibrium direction \cite{matsumoto_voltage-induced_2018, matsumoto_voltage-induced_2019, matsumoto_heavily_2022}.
For the small applied field satisfying $0 < H_{\rm ext} <M_{s} (N_{y} - N_{x})$,
the lower boundary is given by
\begin{equation}
  \label{eq:LB1cond}
  K_{\rm eff,L} = \frac{ \mu_{0} M_{s}^{2} (N_{x} - N_{y})}{2}-\frac{ \mu_{0} M_{s} H_{\rm ext} }{1- m_{y}^{(0)}},
\end{equation}
where
\begin{equation}
  \label{eq:my0A}
  m_{y}^{(0)} = \frac{ \mu_{0} M_{s} H_{\rm ext}}{ 2 K_{u}^{(0)} + \mu_{0} M_{s}^{2} (N_{y}- N_{z}) }.
\end{equation}
The upper boundary is given by \cite{matsumoto_heavily_2022}
\begin{align}
  \label{eq:UB2cond}
  K_{\rm eff,U}
  = & \frac{\mu_{0}} {2 \left[ 1 - \left(m_{y}^{(0)} \right) ^2\right] N_{yx} }
  \Biggl\{
  H_{\rm ext}^2 - 2 M_{s} H_{\rm ext} m_{y}^{(0)} N_{yx}
  \nonumber                                                                                        \\
    & + M_{s}^{2} N_{yx} \left[ N_{z} - N_{x} - \left( m_{y}^{(0)} \right)^2 (N_{z}-N_{y}) \right]
  \Biggr\}
  \nonumber                                                                                        \\
    & - \frac{1}{2} \mu_{0} M_{s}^{2} (N_{z} - N_{x} ) ,
\end{align}
where $N_{yx}=N_{y} - N_{x}$.
For the external magnetic field satisfying $M_{s} (N_{y} - N_{x}) \le H_{\rm ext} < 2 K_{u}^{\rm (0)}/(\mu_{0} M_{s}) + M_{s} (N_{y} - N_{z}) $, the lower boundary is the same as Eq. \eqref{eq:LB1cond} while the upper boundary becomes \cite{matsumoto_heavily_2022}
\begin{equation}
  \label{eq:UB1cond}
  K_{\rm eff,U}
  =
  \frac{ \mu_{0} M_{s} H_{\rm ext} }{ m_{y}^{(0)} + 1}
  - \frac{ \mu_{0} M_{s}^{2} (N_{y} - N_{x}) }{2}.
\end{equation}
Increasing the external magnetic field reduces the energy barrier between the two equilibrium directions. When the external magnetic field is larger than $2 K_{u}^{\rm (0)}/(\mu_{0} M_{s}) + M_{s} (N_{y} - N_{z})$, the energy barrier vanishes and system has one equilibrium direction at $m_{y}^{(0)}=1$. The information cannot be stored as the direction of the magnetization under such a strong magnetic field.

\subsection{$t_{p}$, $K_{\rm eff}^{\rm (+V)}$, and $H_{\rm ext}$ dependence of WER}

\begin{figure}
  \centerline{
    \includegraphics [width=1\columnwidth] {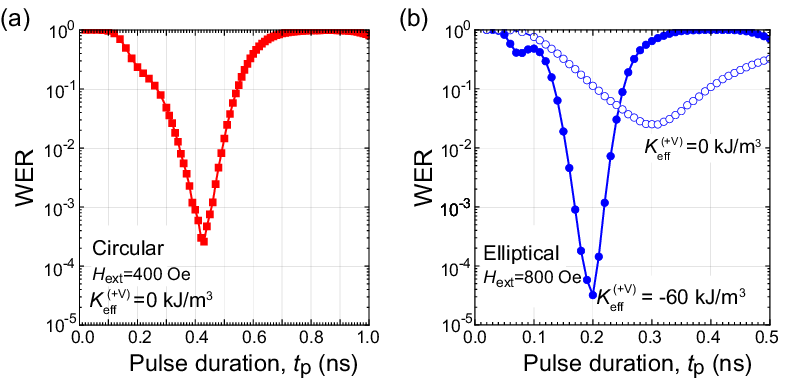}
  }
  \caption{
    \label{fig:wtp}
    (a) Pulse duration, $t_{p}$, dependence of WER for circular MTJ under $H_{\rm ext} = 400$ Oe. The effective out-of-plane anisotropy constant during the pulse is $K_{\rm eff}^{\rm (+V)}=0$ kJ/m$^{3}$. The minimum value of the WER, (WER)$_{\rm min}$, is $2.6\times 10^{-4}$.
    (b) Pulse duration dependence of WER for elliptical MTJ under $H_{\rm ext} = 800$ Oe. The results for $K_{\rm eff}^{\rm (+V)}=0$ and -60 kJ/m$^{3}$ are represented by the open and solid circles, respectively. The minimum value of the WER for $K_{\rm eff}^{\rm (+V)}=0$ and -60 kJ/m$^{3}$ are (WER)$_{\rm min}=2.5\times 10^{-2}$ and $3.2\times 10^{-5}$, respectively.
  }
\end{figure}

Figures \ref{fig:wtp}(a) and \ref{fig:wtp}(b) shows examples of the WER of the circular and elliptical MTJs. 
The thickness and area of the FL is the same for both circular and elliptical MTJs. The demagnetizing coefficient for the circular MTJ is $N_{x}=N_{y}=0.04447$ and $N_{z}=0.91106$ \cite{beleggia_demagnetization_2005}. 
The other parameters, except for $H_{\rm ext}$ and $K_{\rm eff}^{\rm (+V)}$, 
are the same for all calculations. We calculated the WER for wide range of $K_{\rm eff}^{\rm (+V)}$ and $H_{\rm ext}$ and found that the optimal value of ($H_{\rm ext}$, $K_{\rm eff}^{\rm (+V)}$) is (400 Oe, 0 kJ/m$^{3}$) for the circular MTJ and (800 Oe, -60 kJ/m$^{3}$) for the elliptical MTJ.

The WER of the circular MTJ exhibits the minimum value of (WER)$_{\rm min}=2.6\times 10^{-4}$ at $t_{p}=0.43$ ns, 
as shown in Fig. \ref{fig:wtp}(a). 
The effective field during the precession comprises only the external field, 
and the optimal value of the pulse duration is approximately the half of the precession period, 
i.e., $t_{p,{\rm opt}}^{\rm (circle, analytical)}$ = $\pi(1+\alpha^{2})/(\gamma_{0} H_{\rm ext}) = 0.45$ ns.
Note that,
in the circular MTJ with $N_{x} = N_{y}$ and the optimal $K_{\rm eff}^{\rm (+V)}=0$,  
the half of the precession period is $t_{p,{\rm opt}}^{\rm (circle, analytical)}$
regardless of the initial direction of magnetization which is thermally fluctuated.
Therefore, the distribution of the magnetization direction is kept small during the precession, 
and subsequently WER is minimized \cite{shiota_reduction_2017}.
However, in the elliptical MTJ with $N_{x} \neq N_{y}$, this method cannot be applied.

The WER of the elliptical MTJ for $K_{\rm eff}^{\rm (+V)}=0$ exhibits a minimum value of (WER)$_{\rm min}=2.5\times 10^{-2}$ at $t_{p}=0.31$ ns. The WER is lowered by decreasing $K_{\rm eff}^{\rm (+V)}$. In the case of $K_{\rm eff}^{\rm (+V)}=-60$ kJ/m$^{3}$, (WER)$_{\rm min}=3.2\times 10^{-5}$ is obtained at $t_{p}=0.20$ ns. 
This optimal $t_{p}$ is about the half of the precession period, 
and is shorter than $t_{p,{\rm opt}}^{\rm (circle, analytical)}$ 
at $H_{\rm ext}=800$ Oe, 0.23 ns.  
The precession period is shortened by IDF and NOAF.
Note that IDF is induced by the elliptical-cylinder shape, and NOAF by the negative $K_{\rm eff}^{\rm (+V)}$ 
can be obtained by the increase of bias voltage and/or the VCMA effect.
The occurrence of NOAF can be confirmed with the optimal $t_{p}$ 
which is shorter than $t_{p,{\rm opt}}^{\rm (circle, analytical)}$ at the applied $H_{\rm ext}$.

The results show that the appropriate combination of the IDF and the voltage-induced NOAF is effective for reducing the WER.
The mechanism of the reduction of the switching time by IDF and voltage-induced NOAF is further discussed in Sec. \ref{sec:discussions}.

\begin{figure}
  \centerline{
    \includegraphics [width=1.0\columnwidth] {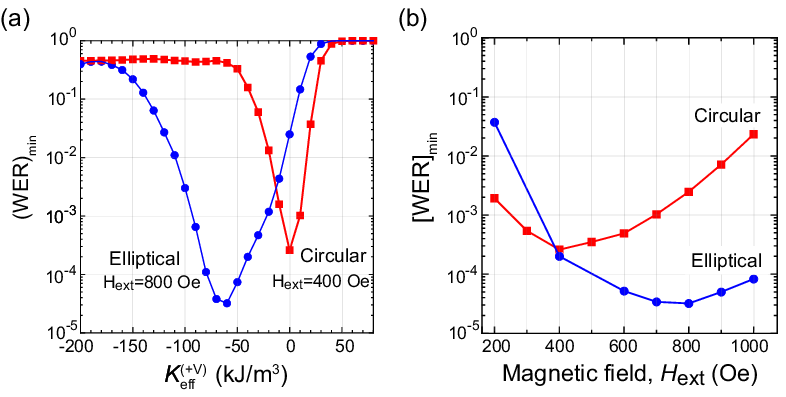}
  }
  \caption{\label{fig:wKH}
    (a) $K_{\rm eff}^{(\rm +V)}$ dependence of (WER)$_{\rm min}$, which is the minimum value of WER in the $t_{p}$ dependence.
    The results for the circular MTJ at $H_{\rm ext}=400$ Oe are represented by red squares connected with red lines. 
    The results for the elliptical MTJ at $H_{\rm ext}=800$ Oe are represented by blue circles connected with blue lines.
    (b) $H_{\rm ext}$ dependence of [WER]$_{\rm min}$ which is the minimum value of (WER)$_{\rm min}$ in the $K_{\rm eff}^{(\rm +V)}$ dependence.
    The results for the circular MTJ 
    are represented by red squares connected with red lines. 
    The results for the elliptical MTJ 
    are represented by blue circles connected with blue lines.
  }
\end{figure}

Figure \ref{fig:wKH}(a) shows the $K_{\rm eff}^{\rm (+V)}$ dependence of (WER)$_{\rm min}$ which is the minimum value of WER obtained by minimizing the WER with respect to $t_{p}$. The red squares connected with the red lines represent the results for the circular MTJ at $H_{\rm ext}=400$ Oe. The blue circles connected with the blue lines represent the results for elliptical MTJ at $H_{\rm ext}=800$ Oe. 
For large positive $K_{\rm eff}^{\rm (+V)}$, the magnetization does not precess, and the WER is approximately unity. 
In the limit of large negative $K_{\rm eff}^{\rm (+V)}$, WER is 0.5. 
This is because the magnetization precesses around $z$-axis during the pulse.
After the pulse, the magnetization relaxes to the two equilibrium directions with equal probability.

The WER is a convex function of $K_{\rm eff}^{\rm (+V)}$, and the lower and upper bounds of $K_{\rm eff}^{\rm (+V)}$ for dynamic switching are given by Eqs. \eqref{eq:LB1cond}, \eqref{eq:UB2cond}, and \eqref{eq:UB1cond}. 
For circular MTJ at $H_{\rm ext}=400$ Oe, $K_{\rm eff,L}^{\rm (c)}=-46.2 $ kJ/m$^{3}$ and $K_{\rm eff,U}^{\rm (c)}=32.5$ kJ/m$^{3}$. 
The lower and upper bounds for dynamic switching of the  elliptical MTJ are $K_{\rm eff,L}=-141$ kJ/m$^{3}$ and $K_{\rm eff,U2}=22.7$ kJ/m$^{3}$, respectively. The region for dynamic switching is spread and shifts to the negative $K_{\rm eff}^{\rm (+V)}$ by the IDF or ellipticity as shown in Fig. \ref{fig:wKH}(a). As a result, the  elliptical MTJ can be used in wider range of $K_{\rm eff}^{\rm (+V)}$ or $V_{p}$ than the circular MTJ.

Figure \ref{fig:wKH}(b) shows $H_{\rm ext}$ dependence of [WER]$_{\rm min}$ which is the minimum value of (WER)$_{\rm min}$ in the $K_{\rm eff}^{(\rm +V)}$ dependence. 
The red squares connected with red lines represent the results for circular MTJ at $K_{\rm eff}^{(\rm +V)}=0$ kJ/m$^{3}$. 
The blue circles connected with blue lines represent the results for elliptical MTJ at $K_{\rm eff}^{(\rm +V)}=-60$ kJ/m$^{3}$. For both MTJs, the [WER]$_{\rm min}$ is a convex function of $H_{\rm ext}$. The decrease of [WER]$_{\rm min}$ with increase of $H_{\rm ext}$ in the small $H_{\rm ext}$ region is due to the decrease of the precession period or the switching time. The increase of the WER with increase of $H_{\rm ext}$ in the large $H_{\rm ext}$ region is due to the decrease of energy barrier between the two equilibrium directions. 
In the absence of the external field, the energy barrier between $(m_{x}$, $m_{y}$, $m_{z}) = (0$, 0, $\pm 1$) and (0, 1, 0) in the elliptical MTJ is higher than that in the circular MTJ because of the in-plane demagnetizing field. 
The enhanced energy barrier in the elliptical MTJ can endure large in-plane $H_{\rm ext}$,
that is, the initial state is stable even at large $H_{\rm ext}$ in the elliptical MTJ \cite{matsumoto_heavily_2022}.
Therefore, the optimal value of $H_{\rm ext}$ that minimize [WER]$_{\rm min}$ of the  elliptical MTJ (800 Oe) is larger than that of the circular MTJ (400 Oe).

In the \ref{sec:appendixA}, 
$\alpha$ and $H_{\rm ext}$ 
dependencies of [WER]$_{\rm min}$ are shown.
In the elliptical MTJ, 
the optimal value of $H_{\rm ext}$ at $\alpha=0.075$ and 0.05
are slightly smaller than that at $\alpha=0.1$, but 
it is larger than that in the circular MTJs.

\section{Discussions}
\label{sec:discussions}
\begin{figure}
  \centerline{
    \includegraphics [width=1\columnwidth] {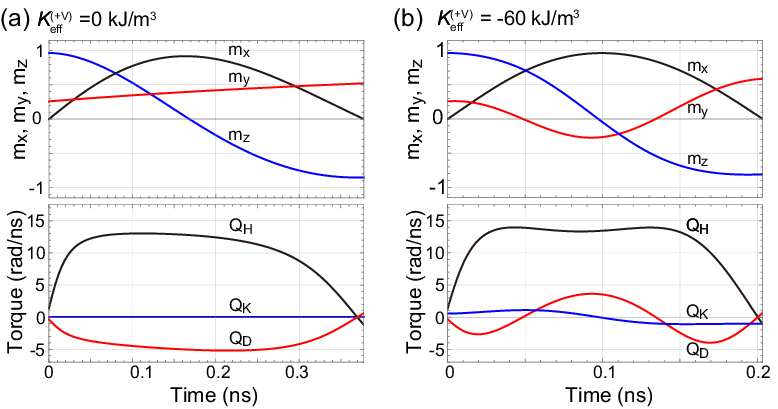}
  }
  \caption{\label{fig:mQ}
    Time evolution of the magnetization and torque of the elliptical MTJ during the pulse at $T=0$ K.
    (a) Results for the pulse yielding $K_{\rm eff}^{\rm (+V)} = 0$ kJ/m$^{3}$. The top panel shows the time evolution of $m_{x}$ (black), $m_{y}$ (red), and $m_{z}$ (blue). The bottom panel
    show the time evolution of the $\theta$ component of the torque due to the external field, $Q_{H}$ (black), the IDF, $Q_{D}$ (red), and the effective out-of-plane anisotropy field, $Q_{K}$ (blue).
    (b) Same plot for the pulse yielding $K_{\rm eff}^{\rm (+V)} = -60$ kJ/m$^{3}$.
  }
\end{figure}

As shown in Figs. \ref{fig:wtp}(a) and (b), the optimal pulse duration of the elliptical MTJ with $K_{\rm eff}^{\rm (+V)} = -60$ kJ/m$^{3}$ (0.20 ns) 
is considerably shorter than the  elliptical MTJ with $K_{\rm eff}^{\rm (+V)} = 0$ kJ/m$^{3}$ (0.31 ns) and the circular MTJ (0.43 ns). The short switching time has the merit of suppressing errors due to the thermal agitation field during the pulse. To understand the mechanism of the reduction of the switching time by IDF and voltage-induced NOAF, 
we plot the time evolution of the magnetization at $T=0$ K and torque of the elliptical MTJ in Figs. \ref{fig:mQ}(a) and \ref{fig:mQ}(b).

The top panel of Fig. \ref{fig:mQ}(a) shows the time evolution of $m_{x}$ (black), $m_{y}$ (red), and $m_{z}$ (blue) of the elliptical MTJ under the pulse yielding $K_{\rm eff}^{\rm (+V)} = 0$ kJ/m$^{3}$. 
The origin of the horizontal axis is set to the beginning of the pulse. Starting from the equilibrium direction with $m_{z} >0$, the magnetization precesses around the external field $H_{\rm ext}$ and reaches the direction with $m_{x}=0$ at 0.38 ns. Due to the damping and the IDF the trajectory is not a semicircle but a semispiral, where $m_{y}$ monotonically increases with time.

In the bottom panel Fig. \ref{fig:mQ}(a), we plot the time evolution of the $\theta$ component of the torque due to the applied field, $Q_{H}$ (black), the IDF, $Q_{D}$ (red), and the effective out-of-plane anisotropy field, $Q_{K}$ (blue). The torques are defined by the equation of motion of $\theta$ as
\begin{align}
  \frac{d\,\theta}{d\tau}
  =
  Q_{H} + Q_{D} + Q_{K},
\end{align}
where
\begin{align}
  Q_{H}
  =
  \frac{H_{\rm ext}}{M_{s}}
  \left(
  \cos\phi + \alpha \cos\theta \sin\phi
  \right),
\end{align}
\begin{align}
  Q_{D}
  =
  \left(N_x - N_y\right)
  \sin\theta \sin\phi
  \left(
  \cos\phi
  +
  \alpha \cos\theta \sin\phi
  \right),
\end{align}
\begin{align}
  Q_{K}
  = -\frac{\alpha\, K_{\rm eff}^{\rm (+V)} \sin 2\theta}{\mu_0 M_{s}^{2}},
\end{align}
and $\tau$ is the dimensionless time defined as
\begin{align}
  \tau
  = \frac{\gamma_{0} M_{s}}{1+\alpha^2}t.
\end{align}
As shown in the bottom panel Fig. \ref{fig:mQ}(a), $Q_{H}$ accelerates the switching dynamics, 
while $Q_{D}$ decelerates the switching dynamics. $Q_{K}$ is zero because $K_{\rm eff}^{\rm (+V)} = 0$ kJ/m$^{3}$.
Note that the deceleration of the switching dynamics by the negative $Q_{D}$ 
increases switching time and therefore WER.

The switching dynamics considerably changes when $K_{\rm eff}^{\rm (+V)}$ is reduced down to -60 kJ/m$^{3}$ as shown in the top panel of Fig. \ref{fig:mQ}(b). The $y$-component of the magnetization, $m_{y}$, oscillates and becomes negative around 0.1 ns. The switching time is as small as 0.20 ns, which is almost half of that for $K_{\rm eff}^{\rm (+V)} = 0$ kJ/m$^{3}$. The reduction of the switching time is caused by the positive $Q_{D}$ around 0.1 ns. 
The results shown in Fig. \ref{fig:mQ}(b) demonstrate that the NOAF induces the rotation of $\bm{m}$ around $z$-axis, 
making $m_{y}$ negative. Then, the in-plane anisotropy field, that is IDF, rotates $\bm{m}$ around the $x$-axis, thereby accelerating the switching dynamics and reducing WER
in the elliptical MTJ. Note that, in the circular MTJ, $Q_{D}(=0)$ shown in Eq. 10 cannot accelerate the switching dynamics 
because $N_{x}=N_{y}$.

\begin{figure}[h]
  \includegraphics [width=1\columnwidth] {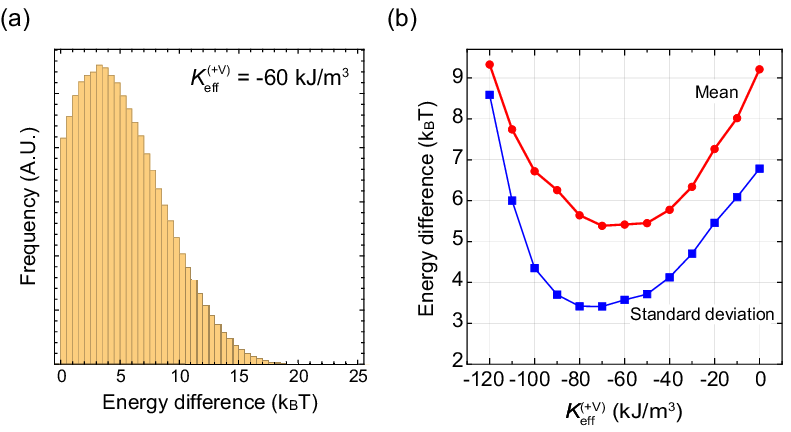}
  \caption{\label{fig:dist}
    (a) Distribution of the energy difference between the magnetization directions at the end of the pulse and the equilibrium direction in the unit of $k_{\rm B}T$. The parameters are assumed to be the optimal values, $K_{\rm eff}^{\rm (+V)}= -60$ kJ/m$^{3}$, $H_{\rm ext}=$ 800 Oe, and $t_{p} = $ 0.20 ns. (b) Mean (red circles) and standard deviation (blue squares) of the energy difference between the magnetization directions at the end of the pulse and the equilibrium direction as a function of $K_{\rm eff}^{\rm (+V)}$.
The external in-plane magnetic field is set as $H_{\rm ext}=$ 800 Oe.    
 At each $K_{\rm eff}^{\rm (+V)}$, $t_{p}$ is set to the optimal values that minimize the WER there.
  }
\end{figure}

Figure \ref{fig:dist}(a) shows the distribution of energy difference between the magnetization directions at the end of the pulse and the equilibrium direction in the unit of $k_{\rm B}T$. The parameters are set to the optimal values, $K_{\rm eff}^{\rm (+V)}= -60$ kJ/m$^{3}$, $H_{\rm ext}=$ 800 Oe, and $t_{p} = $ 0.20 ns. The shift of the peak of the distribution from zero is caused by Gilbert damping. The mean and standard deviation are 5.6 $k_{B} T$ and 3.4 $k_{B}T$, respectively. Figure \ref{fig:dist}(b) shows the $K_{\rm eff}^{\rm (+V)}$ dependence of the mean (red circles) and standard deviation (blue squares) of the energy difference between the magnetization directions at the end of the pulse and the equilibrium direction. 
The external in-plane magnetic field is set as $H_{\rm ext}=$ 800 Oe.
At each $K_{\rm eff}^{\rm (+V)}$, $t_{p}$ is set to the optimal values that minimize the WER there. The mean of the energy difference is minimized around $K_{\rm eff}^{\rm (+V)} = -60$ kJ/m$^{3}$ which is the optimal value at with the WER is minimized. The standard deviation of the energy difference is minimized at $K_{\rm eff}^{\rm (+V)} = -80$ kJ/m$^{3}$.

The standard deviation is a measure of how large the distribution is spread during the pulse and should decrease with decrease of the switching time. Moreover, the mean is a measure of how close the magnetization at the end of the pulse is to the equilibrium direction. The mean depends not only on the switching time but also on the details of torques during the pulse. Therefore, the value of $K_{\rm eff}^{\rm (+V)}$ that minimizes the mean should not be the same as that minimizes the standard deviation. However, the results show that both mean and standard deviation are minimized at $K_{\rm eff}^{\rm (+V)}$ close to the optimal value. The appropriate combination of the IDF and voltage-induced NOAF reduces not only the switching time but also the energy difference between the magnetization directions at the end of the pulse and the equilibrium direction. As a result, the WER is minimized at the optimal value of $t_{p}$, $H_{\rm ext}$, and $K_{\rm eff}^{\rm (+V)}$.

\section{Conclusions}
We theoretically analyzed the WER of voltage-induced switching for the VC-MRAM and showed that the appropriate combination of the in-plane demagnetizing field and voltage-induced negative out-of-plane anisotropy field reduces the WER by one order of magnitude compared with that of the dynamic switching in a conventional MTJ. The mechanism of WER reduction is discussed based on the magnetization dynamics at $T=0$ K and the distribution of the energy difference between the magnetization directions at the end of the pulse and the equilibrium direction at $T=300$ K. 
Furthermore, we show that the reduction of the switching time and the mean of the energy difference by the in-plane demagnetizing field and voltage-induced negative out-of-plane anisotropy field 
causes the reduction of WER. The results provide a guide for designing a reliable VC-MRAM.

\section{Acknowledgements}
This study is partly based on results obtained from a project, JPNP16007,
commissioned by the New Energy and Industrial Technology Development Organization (NEDO), Japan.

\appendix
\section{$H_{\rm ext}$ dependence of the WER}
\label{sec:appendixA}

\begin{figure}
\includegraphics[width=0.7\columnwidth]{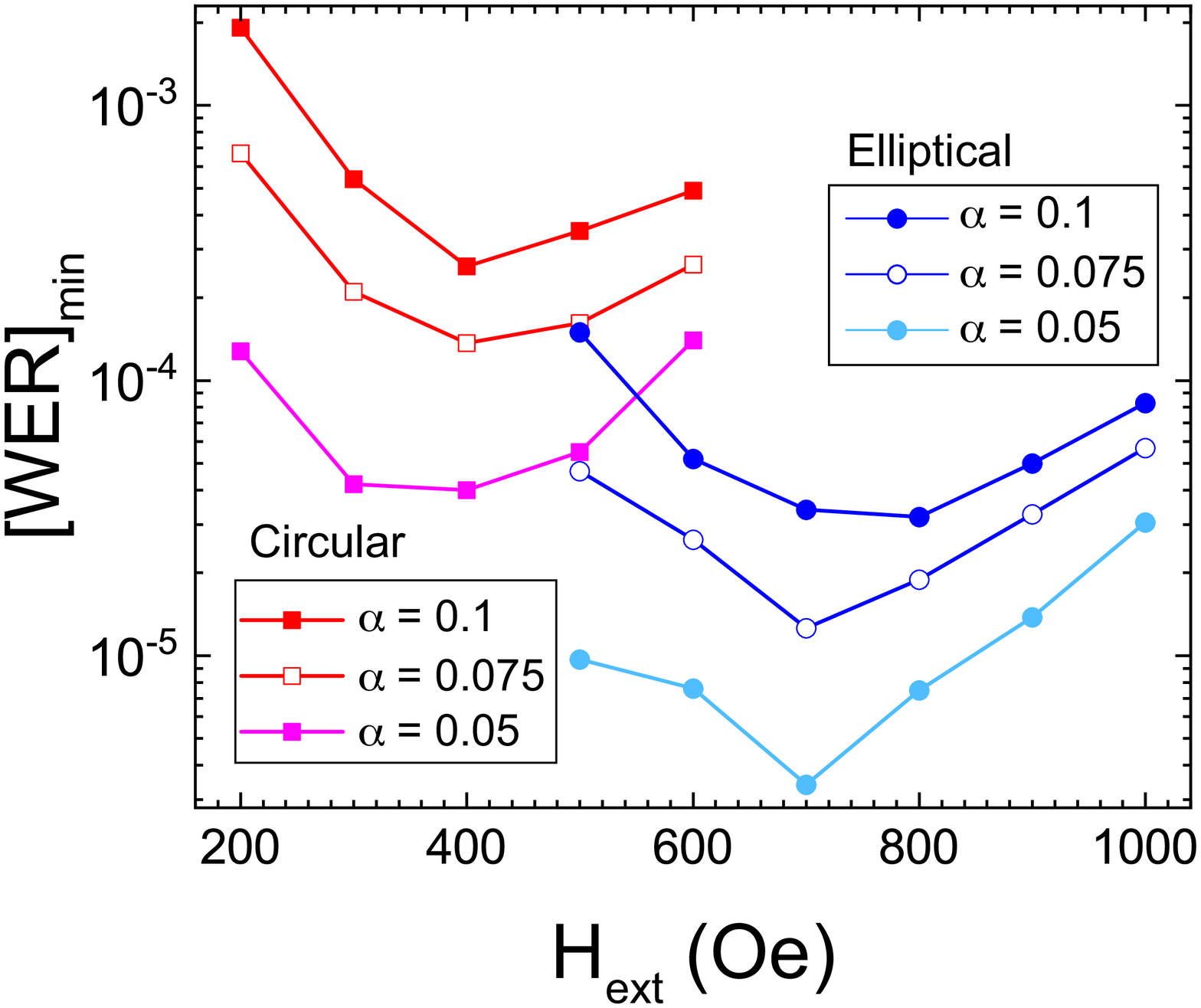}
\caption{\label{fig:fig7} 
    $\alpha$ and $H_{\rm ext}$ dependencies of [WER]$_{\rm min}$.
    The results for the circular MTJ 
    are represented by squares. 
    The results for the elliptical MTJ 
    are represented by the circles.
    The results for the $\alpha=0.1$ are the same as those in 
    Fig. \ref{fig:wKH}(b).
}
\end{figure}

Figure \ref{fig:fig7} shows $\alpha$ and $H_{\rm ext}$ dependencies of [WER]$_{\rm min}$.
For both the circular MTJ and the elliptical MTJ,
[WER]$_{\rm min}$ decreases with decrease of $\alpha$
because the magnitude of the thermal agitation field is proportional to $\sqrt{\alpha}$.

In the circular MTJ with $\alpha=0.1$, 0.075 and 0.05, 
the minimum [WER]$_{\rm min}$ is yielded 
by the optimal conditions, $K_{\rm eff}^{\rm (+V)}=0$ kJ/m$^{3}$ and $H_{\rm ext}=400$ Oe.
In the elliptical MTJ with $\alpha=0.1$, (0.075 and 0.05,) 
the minimum [WER]$_{\rm min}$ is yielded 
by the optimal conditions, $K_{\rm eff}^{\rm (+V)}=-60$ kJ/m$^{3}$, (-60 kJ/m$^{3}$  and -60 kJ/m$^{3}$,)  
and $H_{\rm ext}=800$ Oe, (700 Oe and 700 Oe,) respectively.
The elliptical MTJ with smaller $\alpha$ 
does not require $H_{\rm ext}=800$ Oe 
where the reduction of energy barrier and the stability of the initial states
are more serious than the reduction of the switching time.


\end{document}